\documentstyle[12pt]{article}

\newcommand{\SU}{{\rm SU}}
\newcommand{\cP}{{\cal P}}
\newcommand{\R}{{\bf R}}
\newcommand{\Z}{{\bf Z}}

\newcommand{\eq}{\begin{equation}}
\newcommand{\eqend}{\end{equation}}
\newcommand{\eqa}{\begin{eqnarray}}
\newcommand{\nonueqa}{\begin{eqnarray*}}
\newcommand{\eqaend}{\end{eqnarray}}
\newcommand{\nonueqaend}{\end{eqnarray*}}
\newcommand{\nonu}{\nonumber \\ \nopagebreak}

\begin{document}
\begin{flushright}
April 8, 1998
\end{flushright}
\vspace{.4cm}
\renewcommand{\thefootnote}{\alph{footnote}}

\begin{center}
{\Large \bf   Novel integrable spin-particle models from gauge
theories on a cylinder}\\
\vspace{1 cm}
{\large Jonas Blom and Edwin Langmann}
\vspace{0.3 cm}

{\em Theoretical Physics, Royal Institute of Technology, S-10044
Sweden} 
\end{center}

\begin{abstract}
We find and solve a large class of integrable dynamical systems which
includes Calogero-Sutherland models and various novel generalizations
thereof.  In general they describe $N$ interacting particles moving on
a circle and coupled to an arbitrary number, $m$, of ${\rm su}(N)$
spin degrees of freedom with interactions which depend on arbitrary
real parameters $x_j$, $j=1,2,\ldots,m$.  We derive these models from
$\SU(N)$ Yang-Mills gauge theory coupled to non-dynamic matter and on
spacetime which is a cylinder.  This relation to gauge theories is
used to prove integrability, to construct conservation laws, and solve
these models.

\end{abstract}

Integrable models have always played a central role in classical and
quantum mechanics.  Most prominent examples, like the Kepler problem,
are systems with few ($\leq 3$) degrees of freedom.  An important
exception is a class of integrable $N$-particle models associated with
the names Calogero, Moser and Sutherland \cite{CM,CS} (for review see
\cite{OPcl}).  These are models for identical particles moving on one
dimensional space and interacting via certain repulsive two-body
potentials $v(r)$.  A well-known example is $v(r)\propto g^2 /\sin^2(g
r)$ (which includes $v(r)\propto 1/r^2$ in the limit $g\to 0$), and we
refer to the corresponding model as CS model.  It is known that these
models allow for interesting generalizations which also have dynamic
spin degrees of freedom \cite{W,MP1}.  The CS model and its
generalizations have recently received quite some attention in
different contexts.  Here we only mention their relation to gauged
matrix models \cite{matrix} and gauge theories on a cylinder
\cite{gauge} which will be relevant for us.\footnote{the latter
relation is implicit already in earlier work; see e.g.\ \cite{gauge0}
}

In this article we find and solve a new class of integrable systems
containing the CS model and their spin-generalizations \cite{W} as
limiting cases.  Our method is to extend and exploit the relation of
the CS model to gauge theories on a cylinder, as will be explained in
detail below.  For simplicity our discussion here is restricted to
classical models, and we only consider a special type of
gauge theory.  Our method is simple, and it should be
possible to generalize it in several different directions: We
conjecture that the corresponding quantum models are also integrable,
and that our method to prove integrability should apply to the quantum
case, too.  (In this context it is worth noting that the
quantum-analog of the gauge theory we consider is closely related to
QCD on a cylinder and in the limit where the masses of the quarks
becomes infinite; a good starting point to the literature on this is
Ref.\ \cite{GPSS}).  Moreover, it would be interesting to extend our
method to other gauge theories (e.g.\ with supersymmetry and/or more
general types of matter fields than the one considered here) and thus
try to find and solve other integrable models.

The models we solve are given by a Hamiltonian
\eqa
\label{CMS}
H=\sum_{\alpha=1}^N \frac{(p^\alpha)^2 }{2} + \frac{1}{2}
\sum_{ \stackrel{\alpha,\beta=1}{\alpha\neq \beta} }^N
\sum_{j,k=1}^m
v_{jk} (q^\alpha-q^\beta) \rho_k^{\alpha\beta} \rho_j^{\beta\alpha}
+
\frac{1}{2} \sum_{ \alpha=1}^N
\sum_{j,k=1}^m
c_{jk} \rho_k^{\alpha\alpha} \rho_j^{\alpha\alpha}
\qquad \quad\quad
\eqaend where $q^\alpha$ and $p^\alpha$ are particle coordinates and
momenta with the usual Poisson brackets $ \{q^\alpha,p^\beta\}
= \delta^{\alpha\beta} $ etc., and
$\rho_j^{\alpha\beta}=\overline{\rho_j^{\beta\alpha}}$ are complex valued
${\rm su}(N)$-spins, i.e.\ $\sum_{\alpha=0}^N \rho_j^{\alpha\alpha}=0$ and
\eq
\label{PB}
\{\rho_j^{\alpha\beta},\rho_k^{\alpha'\beta'} \} =
ig 2\pi \delta_{jk}\left(
\delta^{\beta\alpha'}\rho_j^{\alpha\beta'} -
\delta^{\beta'\alpha}\rho_j^{\alpha'\beta} \right) . \eqend
(The other Poisson brackets vanish.) The interaction
potentials are given by
\eq
\label{V}
v_{jk} (r) = \frac{1}{4}e^{-i g r x_{jk} }
\left(\frac{1}{\sin^2(\pi g r)} + \frac{i x_{jk}}{\pi}  \cot(\pi g r)
-\frac{|x_{jk}|}{\pi}\right)
\eqend where
\eq
x_{jk} = (x_j-x_k)_{2\pi}
\eqend
with $s_{2\pi} : = s - 2\pi n$ for the integer $n$ such that
$-\pi\leq s - 2\pi n <\pi$, and
\eq
\label{cjk}
c_{jk} = \frac{x_{jk}^2}{8\pi^2} - \frac{|x_{jk}|}{4\pi} +
\frac{1}{12} .
\eqend
The parameters $g$ (real positive), $N$ and $m$ (positive integers) are
arbitrary, and $-\pi\leq x_1\leq x_2\leq \ldots \leq x_m<\pi$.
Furthermore we have the following constraint on
the possible initial conditions,
\eq
\label{condition}
\sum_{j=1}^m \rho_j^{\alpha\alpha}= 0
\quad \forall \alpha.
\eqend
Note that $v_{jk}(r) = \overline{v_{jk}(-r)}=v_{kj}(-r)$, which implies
that the
Hamiltonian is real. Moreover, since
\eq
v_{jk} (r+\frac{n}{g}) = e^{-in(x_j-x_k)}v_{jk} (r),
\eqend
the Hamiltonian is invariant under the following transformations,
\eq
q^\alpha\to q^\alpha + \frac{n^\alpha}{g},\quad
p^\alpha\to p^\alpha,\quad
\rho_j^{\alpha\beta} \to \rho_j^{\alpha\beta}e^{-i x_j( n^\alpha-n^\beta)}
\eqend
for all integers $n^\alpha$. Thus
these models describe particles
moving on a circle of length $1/g$ and interacting with a potential
whose strength depends on dynamic spins. We note that the particles
repel each other, and we have a further natural
restriction on phase space,
\eq
\label{qq}
q^\alpha \neq q^\beta  \quad \forall \alpha\neq\beta .
\eqend
The main result of this article is a proof of integrability and
the explicit solution of all these models.

Next the relation of our particle-spin models to gauge theories is
discussed.  It is known that the CS model can be obtained from a
gauged one dimensional matrix model \cite{matrix}.  More recently a
relation to gauge theories in 1+1 dimensions was pointed out
\cite{gauge}.  In this article we explore this relation further and
use it to find and solve new integrable models.  We present a simple
argument that $\SU(N)$ Yang-Mills gauge theory on a cylinder coupled
to certain {\em non-dynamic} matter is equivalent to a model of
interacting particles and spins.  We then show that this equivalence
can be used as a powerful tool to analyze and solve these models:
integrability is manifest, the construction of conservation laws
trivial, and a simple solution method is obtained by exploiting gauge
invariance, i.e.\ the possibility to change from the {\em Weyl
gauge}\footnote{$A_0$ and $A_1$ will be defined further below} $A_0=0$
to what we call the {\em diagonal Coulomb gauge}, i.e.\ the condition
that the spatial component of the Yang-Mills field, $A_1(t,x)$, is
independent of $x$ and diagonal in color space, \eq
\label{DCG}
A_1(t,x)= Q(t) = diag\left( q^1(t), q^2(t), \ldots, q^N(t) \right) \:
.  \eqend This is due to the fact that the gauge theory model in the
Weyl gauge is free and can be solved trivially, whereas in the
diagonal Coulomb gauge the time evolution equations are non-linear
and, in a special case, equal to the Hamilton equations of the model
given by Eq.\ (\ref{CMS}).  To be more specific: We restrict ourselves
to gauge theory models with matter fields localized at a finite number
of points $x_j$, $j=1,2,\ldots m$ for simplicity (see Eq.\ (\ref{rho})
below).  We find that the dynamics of such a model in the diagonal
Coulomb gauge is governed by the equations of motion $\dot X= \{ X,H
\}$, $X=q^\alpha,p^\alpha,\rho_j^{\alpha\beta}$, which follow from the
Hamiltonian Eq.\ (\ref{CMS}) and the Poisson brackets $\{\cdot,\cdot\} $
given above.  These observations allows us to derive the full solution
of the initial value problem for all these models generalizing the
known solution of the CS model \cite{OPcl}.

We now consider 1+1 dimensional Yang-Mills theory, i.e.\ the
differential equations $\sum_{\mu=0,1}[D_\mu,F^{\mu\nu}] = J^\nu$
where $D_\mu = \partial_\mu + ig A_\mu$, $A_\mu$ are the Yang-Mills
fields, $J^\mu$ the matter currents, $g$ the Yang-Mills coupling
strength, $\partial_\mu= \partial/\partial x^\mu$, and $\mu,\nu=0,1$.
Spacetime is a cylinder i.e.\ $t=x^0\in\R$ is time, and $x=x^1\in
[-\pi,\pi]$ (= circle).  Moreover,\footnote{$[a,b]:=ab-ba$}
$F_{\mu\nu} = [D_\mu,D_\nu]/ig $, and our metric tensor is
$diag(1,-1)$.  As gauge group we take $\SU(N)$ in the fundamental
representation.\footnote{i.e.\ $A_\mu$, $E$ and $\rho$ are
functions with values in the traceless, complex $N\times N$
matrices}

We restrict ourselves to {\em non-dynamic matter}, i.e.\ $J^1=0$, and
$J^0\equiv\rho$. We denote $\rho$ as {\em charge}.
Note that we have to impose $[D_0,\rho]=0$ for consistency.

Setting $E:=\, F_{01}$, we can write these equations as
follows,
\eqa
\partial_0 A_1 = E + \partial_1 A_0 + ig  [A_1,A_0] \label{1} \\
\partial_0 E + ig  [A_0,E] = 0 \label{2} \\
\partial_0 \rho + ig  [A_0,\rho] = 0 \label{3}\\
\partial_1 E + ig  [A_1,E] = \rho \: . \label{4}
\eqaend
Eq.\ (\ref{4}) is called Gauss' law and is a constraint on
possible initial data for the system of time evolution equations
(\ref{1})--(\ref{3}). We now exploit gauge invariance:
Eqs.\ (\ref{1})--(\ref{4}) are obviously invariant under gauge transformations
\eqa
\label{gt}
A_\mu &\to&  U^{-1} A_\mu U + \frac{1}{ig } U^{-1}\partial_\mu U \nonu
E &\to&U^{-1} E U \nonu
\rho &\to& U^{-1} \rho U
\eqaend
where $U=U(t,x)$ is an arbitrary differentiable $\SU(N)$-valued function on
spacetime.  To eliminate the gauge degrees of freedom one has to fix a
gauge.  One convenient choice is the Weyl gauge
$A_0(t,x)=0$.\footnote{i.e.\ to consider the model in terms of
the gauge transformed fields on the r.h.s. of Eq.\ (\ref{gt}), which by
abuse of notation we denote by the same symbol, and
with a gauge transformation $U$  which is a solution of
$\partial_0 U + ig A_0 U  = 0$. }
Then the Eqs.\ (\ref{1})--(\ref{3}) can be solved trivially:
\eqa
\label{A0}
E(t,x)&=& E(0,x) , \nonu
A_1(t,x) &=& A_1(0,x) + E(0,x) t , \nonu
\rho(t,x) &=& \rho(0,x)
\eqaend
with the initial data $E(0,x)$, $A_1(0,x)$ and $\rho(0,x)$ satisfying the
Gauss' law Eq.\ (\ref{4}) (note that our solution Eq.\ (\ref{A0}) satisfies
Eq.\
(\ref{4}) for all $t$ if it satisfies it for $t=0$).

As mentioned, $E,A_\mu$ and $\rho$ are functions with values in the
traceless $N\times N$ matrices.  In the following we write the matrix
elements of $M=E,A_\mu$ or $\rho$ as $ M^{\alpha\beta}$,
$\alpha,\beta=1,2,\ldots N$.  Note that, since $\sum_{\alpha=0}^N
M^{\alpha\alpha}$ is zero, the independent components are
$M^{\alpha\beta}$ for $\alpha\neq\beta$, and $ M^{\alpha\alpha} -
M^{\alpha+1,\alpha+1} $ for $\alpha=1,2,\ldots N-1$.

We now show that one can also impose the diagonal Coulomb gauge
(\ref{DCG}), i.e.\ for each (generic) Yang-Mills configurations
$A_1(t,x)$ one can find a gauge transformation $U$ such that
$A_1^U\equiv U^{-1} A_1 U + U^{-1}\partial_1 U/i g$ is a diagonal
matrix $Q$ independent of $x$ \cite{LS}.  For that we construct such a
$U$ explicitly.  We first note that a solution to the equation
$\partial_1 S +ig A_1 S=0$ with $S(t,-\pi)=1$ is the parallel
transporter \eq
\label{S}
S(t,x)= \cP \exp\left( -ig \int_{-\pi}^x dy \, A_1(t,y) \right)
\eqend
where $\cP\exp$ is the path ordered exponential.  Note that $S(t,x)$ is not a
gauge transformation since it is not periodic in $x$ (its values at
$x=-\pi$ and $\pi$ are different in general).  To construct a
gauge transformation, we introduce the $\SU(N)$-matrix $V(t)$
diagonalizing the $\SU(N)$-matrix $S(t,\pi)$,
\eq
\label{VSV}
V(t)^{-1} S(t,\pi) V(t) = e^{-ig  2\pi  Q(t) }
\eqend
for some diagonal matrix  $Q(t)$.  This implies that
\eq
\label{U}
U(t,x) = S(t,x) V(t) e^{ig (x+\pi) Q(t) } \eqend is periodic in
$x$, and it satisfies $\partial_1 U +ig A_1 U= ig U Q $
equivalent to $A_1^U=Q$.  Moreover, if $A_1(t,x)$ is a {\em generic}
differentiable map on spacetime, then $Q(t)$ and $V(t)$ can be
chosen to be differentiable in $t$ \cite{Blau}, and $U(t,x)$ Eq.\
(\ref{U})
is indeed a differentiable function on space-time i.e.\ a
gauge transformation.  `Generic' here means that the latter is only
true if $q^\alpha(t)\neq q^{\beta}(t)$ for all $t$ and
$\alpha\neq\beta$ since otherwise discontinuous functions $V(t)$ can
occur \cite{Blau}.  Due to Eq.\ (\ref{qq}), gauge field configurations
$A_1(t,x)$ where this condition fails are irrelevant for us.  Note
that our discussion here implies that
the $q^\alpha(t)$ can be
obtained as eigenvalues of the Wilson line $S(t,\pi)$.  This
observation will allow us to determine the explicit solution of the
Hamilton eqs.\ following from Eqs.\ (\ref{CMS}) and (\ref{V}).

We now determine the time evolution equations for the variables
$q^{\alpha}$ defined in Eq.\ (\ref{DCG}).  We use Fourier transformation,
\eqa
\hat E^{\alpha\beta} (t,n) = \int_{-\pi}^{\pi} dx\, e^{-i nx}
E^{\alpha\beta} (t,x), \quad n\in\Z
\eqaend
and similarly for $A_0$ and $\rho$. Then Eq.\ (\ref{1}) gives
\eq
\label{qdot}
\partial_0 q^{\alpha}(t) = p^{\alpha}(t) \equiv
\frac{\hat E^{\alpha\alpha} (t,0)}{2\pi} \: .
\eqend
Note that this and the following equations all are consistent
with $\sum_{\alpha=1}^N q^{\alpha}=\sum_{\alpha=1}^N p^{\alpha}=0$
(this corresponds to translation invariance of the mechanical system
defined in Eq.\ (\ref{CMS})). The time evolution of the $p^\alpha$ follows
from Eq.\ (\ref{2}),
\eqa
\partial_0 p^{\alpha}(t) = -\frac{ig }{(2\pi)^2}\sum_{n\in\Z}
\sum_{\stackrel{\beta=1}{\beta\neq\alpha}}^N \left(
\hat A_0^{\alpha\beta}(t,n) \hat E^{\beta\alpha}(t,-n) -
\right.\nonu\left.
- \hat E^{\alpha\beta}(t,n) \hat A_0^{\beta\alpha}(t,-n) \right)
\: .
\eqaend
The r.h.s. of this equation can be evaluated using Eqs.\
(\ref{1}) and (\ref{4})
\eqa
\label{EA0}
 -i \left(n + g [q^{\alpha}(t)-q^{\beta}(t) ]\right) \hat
 A_0^{\alpha\beta} (t,n) =
\hat E^{\alpha\beta} (t,n) \nonu
i \left(n + g [q^{\alpha}(t)-q^{\beta}(t) ]\right) \hat
E^{\alpha\beta} (t,n) =\hat\rho^{\alpha\beta} (t,n)
\eqaend
(note that this holds true even for $\alpha=\beta$ if $n\neq 0$).
Inserting this we get \eqa
\label{Pdot}
\partial_0 p^{\alpha}(t) = \frac{2 g }{(2\pi)^2}\sum_{n\in\Z}
\sum_{\stackrel{\beta=1}{\beta\neq\alpha}}^N
\frac{\hat \rho^{\alpha\beta}(t,n) \hat \rho^{\beta\alpha}(t,-n)}
{ \left(n +  g [q^{\alpha}(t)-q^{\beta}(t) ]\right)^3} \: .
\eqaend

Note that up to now no specific choice for the charges was made. To
proceed, we restrict ourselves to charges of the following form for
simplicity,
\eqa
\label{rho}
\rho^{\alpha\beta}(t,x) \equiv \sum_{j=1}^m
\rho^{\alpha\beta}_{j}(t)\delta(x-x_j),
\eqaend
which describe matter localized at the points $x_j$, as mentioned above.
Then
\eq
\label{rhohat}
\hat\rho^{\alpha\beta}(t,n)=\sum_{k=1}^m \rho_k^{\alpha\beta}(t)
e^{-in x_k} \: ,
\eqend
and we can then write Eq.\ (\ref{Pdot}) as
\eq
\label{pdot}
\partial_0 p^\alpha (t)  = - \frac{\partial}{\partial q^\alpha(t) }
\sum_{ \stackrel{\beta=1}{\beta\neq\alpha} }^N\sum_{j,k=1}^m
\rho_k^{\alpha\beta}(t) \rho_j^{\beta\alpha}(t)
v_{jk} (q^\alpha(t)-q^\beta(t) )
\eqend
with
\eq
\label{V1}
v_{jk}(r) =
\sum_{n}
\frac{e^{i n(x_j-x_k) } }{(2\pi)^2(n+g r)^2}
\eqend
where the summation is over all $n\in\Z$.
This equals Eq.\ (\ref{V}), as can be seen by a simple computation
using the identity
\eq
\label{id}
\sum_{n\in\Z}\frac{e^{ins} }{(n+r)^2} =
e^{-i rs_{2\pi} }\left(\frac{\pi^2}{\sin^2(\pi r)}
+ i \pi s_{2\pi}  \cot(\pi r) -\pi|s_{2\pi}|
\right) .
\eqend
Note also that
\eq
\label{id1}
\sum_{n\neq 0}\frac{e^{ins} }{n^2} = \frac{s_{2\pi}^2}{2}-\pi|s_{2\pi}| +
\frac{\pi^2}{3}.
\eqend
Eqs. (\ref{qdot}) and (\ref{pdot}) are precisely the Hamilton equations
$\dot q^\alpha = \{ q^\alpha,H\}$ and $\dot p^\alpha = \{ p^\alpha,H\}$.
We are left to determine the time evolution of the $\rho^{\alpha\beta}_j$.
From Eq.\ (\ref{3}) we get
\eq
\label{Sdot}
\partial_0 \rho^{\alpha\beta}_j(t) = -ig \sum_{\gamma=1}^N \left(
A_0^{\alpha\gamma}(t,x_j)\rho^{\gamma\beta}_j(t) -
\rho^{\alpha\gamma}_j(t)A_0^{\gamma\beta}(t,x_j) \right) .  \eqend
Moreover, we can compute $A_0^{\alpha\beta}(t,x_j)=\frac{1}{2\pi}
\sum_n \hat A_0^{\alpha\beta}(t,n) e^{in x_j}$ from Eqs.\ (\ref{EA0})
and Eq.\ (\ref{rhohat}).  Note that Eq.\ (\ref{EA0}) also determines
$\hat A_0^{\alpha\alpha}(t,n)$ if $n$ is non-zero, and we can set
$\hat A_0^{\alpha\alpha}(t,0)=0$.\footnote{A non-zero $\hat
A_0^{\alpha\alpha}(t)=0$ is irrelevant since it can be removed by a
gauge transformation compatible with the DCG.} Then a simple
computation gives \eq
\label{A0p}
A_0^{\alpha\beta}(t,x_j) = 2\pi \sum_{k=1}^m
v_{jk}(q^\alpha(t)-q^\beta(t)) \rho^{\alpha\beta}_k(t) \eqend with
$v_{jk}(r)$ given by Eq.\ (\ref{V}) and $v_{jk}(0)=c_{jk}$ by Eq.\
(\ref{cjk}).  (We used Eqs.\ (\ref{V1}), (\ref{id}) and (\ref{id1})).
Note that for $\alpha=\beta$ the summation in Eq.\ (\ref{V1}) is
restricted to the non-zero integers $n$, which implies
$v_{jk}(0)=c_{jk}$.)  With that Eq.\ (\ref{Sdot}) becomes equal to the
Hamilton eq.\ $\dot \rho^{\alpha\beta}_j = \{ \rho^{\alpha\beta}_j ,H
\}$ following from Eqs.  (\ref{CMS}) and (\ref{PB}).

We finally note that the $n=0$ components of Gauss' law Eq.\ (\ref{4}) for
$\alpha=\beta$ reads
\eq
\hat\rho^{\alpha\alpha}(t,0) = 0\quad \forall \alpha \: .
\eqend
This is a consistency requirement.  Our arguments above show that this
condition is fulfilled for $\rho$ in Eq.\ (\ref{rho}) if and only if Eq.\
(\ref{condition}) holds for all $t$, which is true if it holds for $t=0$.

We now solve the Eqs.\ (\ref{qdot}), (\ref{pdot}) and
(\ref{Sdot})--(\ref{A0p})
with the initial conditions
\eqa
\label{IC}
q^{\alpha}(0)=q^\alpha_0,\quad  p^{\alpha}(0)=p^\alpha_0,\quad
\rho^{\alpha\beta}_j(0)=\rho^{\alpha\beta}_{j,0} .
\eqaend
Our discussion above implies that we can obtain the solution
$q^{\alpha}(t)$ of this initial value problem by solving the Yang-Mills
Eqs.\ (\ref{1})--(\ref{4}) with the initial conditions
\eqa
\label{ICYM}
A_1^{\alpha\beta}(t=0,x) = \delta_{\alpha\beta}q^\alpha_0,\quad
\int_{-\pi}^{\pi} \frac{dx}{2\pi} E^{\alpha\alpha}(t=0,x) = p^\alpha_0
\nonu \rho^{\alpha\beta}(t=0,x) = \sum_{j=1}^N
\rho^{\alpha\beta}_{j,0}\delta(x-x_j) \: .
\eqaend
We first have to determine $E^{\alpha\beta}(t=0,x)$ for $\alpha\neq\beta$
from Gauss' law Eq.\ (\ref{4}).  The solution $A_1(t,x)$ of the gauge theory
in the Weyl gauge $A_0=0$ is then given in Eq.\ (\ref{A0}).  To obtain the
$q^\alpha(t)$, we only need to evaluate the corresponding parallel
transporter $S(t,\pi)$ Eq.\ (\ref{S}): as discussed, the eigenvalues
of $S(t,\pi)$ are equal to $e^{-2\pi ig q^{\alpha}(t)}$. Moreover,
\eq
\label{Sjt}
\rho_j(t)=U(t,x_j)^{-1} \rho_{j,0}U(t,x_j)
\eqend
with $U(t,x)$ given by Eq.\ (\ref{U}). Here and in the following we use an
obvious
matrix notation.

For $t=0$ we can write Gauss' law Eq.\ (\ref{4}) as
follows\footnote{Note that the same argument applies for
times $t>0$}
\eq
\label{22}
\partial_1 \left( e^{ig Q_0 x} E(0,x) e^{-ig Q_0x} \right)  =
e^{ig Q_0x}\rho(0,x) e^{-ig Q_0x}
\eqend
with $\rho(0,x)=\sum_{j=1}^m \rho_{j,0}\delta(x-x_j)$ and
\eq
\label{Q0}
Q_0=diag\left( q^1_0,q^2_0,\ldots q^N_0\right)\: .
\eqend
Since $\rho(0,x)=0$ except for $x=x_j$, we obtain $E(0,x)= e^{-ig Q_0
x}B_j e^{ig Q_0 x}$, where $B_j$ is some
constant matrix, for $x_j<x<x_{j+1}$, $j=0,\ldots,m$, $x_0=-\pi$ and
$x_{m+1}=\pi$.  To determine the matrices $B_j$ we integrate Eq.\
(\ref{22}) from $x_j-0^+$ to $x_j+0^+$.  This gives the recursion
relations $ B_{j} - B_{j-1} = e^{ig Q_0x_j}\rho_{j,0} e^{-ig
Q_0 x_j}$, and the condition $E(0,-\pi)=E(0,\pi)$ implies $e^{
2ig Q_0\pi}B_0 e^{-2ig Q_0\pi } = B_m$.  Putting this together
and using the second relation in Eq.\ (\ref{ICYM}) we obtain after a
straightforward calculation, \eqa
\label{B}
B_j^{\alpha\beta} = \delta_{\alpha\beta}\left( p_0^{\alpha}
+\sum_{\ell = 1}^j\rho^{\alpha\alpha}_{\ell,0} -
\sum_{i = 1}^m \frac{x_{i+1}-x_i}{2\pi}\sum_{\ell=1}^i
\rho^{\alpha\alpha}_{\ell,0}
\right) +\nonu
\left(1-\delta_{\alpha\beta}\right) \sum_{\ell=1}^m \frac{
\rho_{\ell,0}^{\alpha\beta}
e^{ ig [q_0^{\alpha}-q_0^{\beta}][x_\ell + \pi{\rm sgn}(x_j-x_\ell)] }  }
{2i\sin(g\pi[q_0^{\alpha}-q_0^{\beta}] )}
\eqaend
with ${\rm sgn}(x)=1$ for $x\geq 0$ and $-1$ for $x<0$.
With that, Eq.\ (\ref{A0}) gives $ A_1(t,x) = e^{-ig Q_0 x} \left( Q_0 +
B_j t\right) e^{ig Q_0 x}$ for $x_j<x<x_{j+1}$. This is the solution
of the Yang-Mills equations in the Weyl gauge.

We now solve $\partial_1 S(t,x)+ig A_1(t,x) S(t,x)=0$ which is
equivalent to
\eq \partial_1 \tilde S(t,x) + ig B_j t\tilde S(t,x)=0 \quad \mbox{
for $x_j<x<x_{j+1}$} \eqend for $\tilde S(t,x) = e^{ig Q_0 x}
S(t,x)$.  This implies $\tilde S(t,x)= e^{-i g B_j
t(x-x_j)}\tilde S(t,x_j)$ for $x_j<x<x_{j+1}$, thus \eqa
\label{Sj}
 S(t,x_{j+1}) = e^{-ig Q_0 x_{j+1} }
e^{-ig B_j t(x_{j+1}-x_j)}\tilde S(t,x_j )
= \ldots = \nonu
= e^{-ig Q_0 x_{j+1} }
e^{-ig B_j t(x_{j+1}-x_j) }e^{-ig B_{j-1} t (x_j-x_{j-1})}
\cdots
\nonu
\times \cdots e^{-ig B_0 t(x_1+\pi)}  e^{ -ig Q_0 \pi}
\eqaend
where we used $S(t,-\pi)={\bf 1}$.  Especially (for $j=m$),
\eqa
\label{W}
S(t,\pi) =  e^{-ig Q_0 \pi}
e^{-ig B_m t(\pi-x_m)} \cdots \nonu \times
e^{-ig B_{m-1} t(x_{m}-x_{m-1})}\cdots
e^{-ig B_{0} t(x_{1}+\pi)}
e^{ -ig Q_0 \pi}    \: .
\eqaend

We thus obtain our main\\

\noindent{\bf Result:} {\em The Hamiltonian equations of the the
dynamical system defined in Eqs.\ (\ref{CMS})--(\ref{condition}) are given
in Eqs.\ (\ref{qdot}), (\ref{pdot}) and (\ref{Sdot}).  The solutions of
these equations with the initial conditions Eq.\ (\ref{IC}) can be
obtained from the eigenvalues of the matrix $S(t,\pi)$ given in Eqs.\
(\ref{W}) and (\ref{B}) according to Eq.\ (\ref{VSV}).  Moreover, $\rho_j(t)$
is given by Eq.\ (\ref{Sjt}) with $U(t,x_j)$ defined in Eq.\ (\ref{U}) and
$S(t,x_j)$ in Eq.  (\ref{Sj}).  }\\

Note that for $m=1$, the dynamics of the spin and the particles
decouple, and our result reduces to the known solution of the CS model;
see e.g.\ \cite{OPcl}.

It is now also easy to construct conservation laws for our dynamical
systems: Eq.\ (\ref{A0}) implies that $tr[E(t,x)^n]$, where $tr$ is
the $N\times N$ matrix trace, is time independent for all $-\pi\leq
x<\pi$ and all positive integers $n$.  Since these quantities are
gauge independent, they are time independent also in the diagonal
Coulomb gauge.  In this latter gauge, we can evaluate $E(t,x)$ as
above and obtain $E(t,x)= e^{-ig Q(t) x}B_j(t)e^{ig Q(t) x}$ for
$x_j<x<x_{j+1}$ where $Q(t)$ and $B_j(t)$ are as in Eqs.\ (\ref{Q0})
and (\ref{B}) but with $q^\alpha_0$, $p_0^\alpha$, and
$\rho_{j,0}^{\alpha\beta}$ replaced by $q^\alpha(t)$, $p^\alpha(t)$,
and $\rho_{j}^{\alpha\beta}(t)$, i.e.\ the solution of the initial
value problem which we solved above.  Using cyclicity of the trace, we
conclude that $tr[B_j(t)^n] $ for an arbitrary positive integer $n$
and $j=1,\ldots m$ are time independent: Each of them is a
conservation law.  For $m=1$ these are the known conservation laws for
the CS model \cite{OPcl}.  It is also worth noting the corresponding
Lax-type equations $\partial_0 B_j(t) + i g [ M_j(t) , B_j(t)] = 0$
where \eq M_j(t) = e^{ig Q(t) x_j}A_0(t,x_j)e^{-ig Q(t) x_j} - x_j
P(t) \eqend which are obtained from eq.\ (\ref{2}) setting $x=x_j$ and
using $\partial_0 Q(t) = P(t): =diag(p_1(t),p_2(t),\ldots,p_N(t))$
($A_0(t,x_j)$ is given by Eq.\ (\ref{A0p})).

As discussed above, the models we find generalize the CS models with
the interaction potential $v(r)\propto g^2 /\sin^2(g r)$ which
describe particle moving on a circle of length $1/g$.  There is a an
integrable CS-type model of particles moving on the real line and
interacting with a potential $v(r)\propto g^2 /\sinh^2(g r)$, see
e.g.\ \cite{OPcl}.  The $\sinh$-model and its solution can be obtained
from the $\sin$-model and its solution by replacing $q^\alpha\to i
q^\alpha$ and $p^\alpha\to i p^\alpha$ \cite{OPcl}.  This replacement
in our Hamiltonian, Eq.\ (\ref{CMS}) (together with $H\to -H$), leads
to spin-generalizations of the $\sinh$-model.  It is natural to
conjecture that this very replacement allows to obtain the solution of
the latter from our solution of the former model.

\vspace*{2mm} \noindent{\bf Acknowledgment:} We would like to thank
A.\ Polychronakos for pointing out Ref.\ \cite{W} to us.  His
explanation of the results of these papers helped us to resolve a
difficulty which had prevented us from completing this work earlier.
We also thank him for useful comments on the manuscript.

\end{document}